\documentclass{PoS}

\title{Staggered Diquarks for Singly Heavy Baryons}

\ShortTitle{Staggered Diquarks for Singly Heavy Baryons}

\author{Steven Gottlieb
        \\
        Department of Physics, Indiana University, Bloomington, 47405, IN, U.S.A.\\
        E-mail: \email{sg@denali.physics.indiana.edu}}

\author{Heechang Na         
        \\
        Department of Physics, Indiana University, Bloomington, 47405, IN, U.S.A.\\
        E-mail: \email{heena@indiana.edu}x
}

\author{\speaker{Kazuhiro Nagata}%
        \\
        Department of Physics, Indiana University, Bloomington, 47405, IN, U.S.A.\\
        E-mail: \email{knagata@indiana.edu}}

\abstract{
In the staggered fermion formulation of lattice QCD,
we construct diquark operators which are to be embedded in 
singly heavy baryons.  
The group theoretical connections between 
continuum and lattice staggered diquark representations
are established.
}

\FullConference{The XXV International Symposium on Lattice Field Theory\\
                 July 30 - August 4 2007\\
                 Regensburg, Germany}

\begin{document}

\section{Introduction}

There have been a number of attempts to investigate 
heavy baryons in terms of experimental as well as theoretical methods. 
In lattice QCD, several calculations have been performed in the
quenched regime \cite{Bowler,Flynn,Mathur,Woloshyn,G-T,Khan,Chiu} 
and given a fair agreement with the experimental results.
In \cite{N-G}, two of the authors have reported the results
of preliminary study for the singly charmed baryon mass spectrum 
using the data of $2+1$ flavors dynamical improved staggered quarks. 
Since the staggered fermion provides very fast simulations and
much less statistical errors compared to the other available frameworks, 
(see, for example, Ref \cite{MILC}),
it is worthwhile to pursue more extensive studies of
those baryons in terms of staggered light quarks.
It is then desired to establish the 
group theoretical classification
of pairs of staggered quarks (staggered diquarks)
in order to extract the desired spin and parity state
in the continuum limit.
In \cite{GNN}, we construct all the possible time-local 
staggered diquarks embedded in singly heavy baryons
and establish the group theoretical connections 
between lattice operators and continuum representations w.r.t.
spin, taste and $2+1$ flavor symmetries.

\section{Staggered Diquarks in the Continuum Spacetime}

A singly heavy baryon operator consists of two light quarks
(up, down or strange) and one heavy quark (charm or bottom (or top)). 
The quantum numbers of singly heavy baryons are listed in Table \ref{baryons}.
In this section, we classify the irreps
of the staggered diquarks 
w.r.t. spin, flavor and taste symmetry group 
in the continuum spacetime.
We especially take $2+1$ as the flavor symmetry group under which 
the recent dynamical simulations of staggered lattice QCD are performed.
\begin{table}
\begin{center}
\renewcommand{\arraystretch}{1.2}
\renewcommand{\tabcolsep}{5pt}
\begin{tabular}{c|cccc||c}
\hline
Baryon & $J^{P}$ &  $Z$ & Content & $(SU(2)_{S},SU(2)_{I})_{Z}$ 
& ($SU(2)_{S}, SU(8)_{x,y}, SU(4)_{z})_{Z}$\\
\hline
$\Lambda_{Q}$ & $\frac{1}{2}^{+}$ & $0$ & $(ll)Q$ 
& $(\bf{1}_{A},\bf{1}_{A})_{\rm 0}$ &
$({\bf{1}_{A}}, {\bf{28}_{A}},{\mathbf{1}})_{0}$ \\
$\Xi_{Q}$ & $\frac{1}{2}^{+}$ & $-1$ &$(ls)Q$  
& $(\bf{1}_{A},\bf{2})_{\rm -1}$ & 
$({\bf{1}_{A}}, {\bf{8}},{\mathbf{4}})_{-1}$ \\
$\Sigma_{Q}^{(*)}$ & $\frac{1}{2}^{+} (\frac{3}{2}^{+})$ & $0$ &$(ll)Q$ 
& $(\bf{3}_{S},\bf{3}_{S})_{\rm{0}}$ &
$({\bf{3}_{S}}, {\bf{36}_{S}},{\mathbf{1}})_{0}$ \\
$\Xi'^{(*)}_{Q}$ & $\frac{1}{2}^{+}(\frac{3}{2}^{+})$ & $-1$ & $(ls)Q$ 
& $(\bf{3}_{S},\bf{2})_{\rm{-1}}$ &
$({\bf{3}_{S}}, {\bf{8}},{\bf{4}})_{-1}$ \\
$\Omega^{(*)}_{Q}$ & $\frac{1}{2}^{+}(\frac{3}{2}^{+})$ & $-2$ & $(ss)Q$ 
& $(\bf{3}_{S},\bf{1})_{\rm{-2}}$ &
$({\bf{3}_{S}}, {\bf{1}},{\bf{10}_{S}})_{-2}$ \\
\hline
\end{tabular}
\caption{Quantum numbers for singly heavy baryons for
$2+1$ flavor symmetry :
$Z$ denotes strangeness.
The states with asterisks represent the spin $\frac{3}{2}$ states.
The fifth and sixth column represent the light diquark irreps
for single taste and four tastes, respectively.}
\label{baryons}
\end{center}
\end{table}

Let us begin with reviewing the 
diquark irreps for mass degenerate light quarks of single taste,
namely physical valence quarks.
The success of non-relativistic $SU(6)$ quark model suggests that
the diquarks 
should belong to the irreps $\bf{21}_{S}$, 
the symmetric part of $\bf{6}\otimes\bf{6}$,
which has the following decomposition into $SU(2)_{S} \times SU(3)_{F}$,
the direct product of non-relativistic spin and $SU(3)$ flavor,
\begin{eqnarray}
SU(6) &\supset& SU(2)_{S} \times SU(3)_{F}  \nonumber \\
\bf{21}_{S} &\rightarrow& (\bf{3}_{S},\bf{6}_{S}) \oplus (\bf{1}_{A},\bf{3}_{A}),
\end{eqnarray} 
where the labeling represents the dimension of each irreps while
the subscripts $\bf_{S}$ and $\bf_{A}$ indicate the symmetric 
and anti-symmetric part, respectively. 
For $2+1$ flavors, $SU(3)_{F}$ is decomposed into $SU(2)_{I}$ isospin group.
Accordingly, we have
\begin{eqnarray}
SU(2)_{S} \times SU(3)_{F} &\supset& SU(2)_{S} \times SU(2)_{I}  \nonumber \\
(\bf{3}_{S},\bf{6}_{S}) &\rightarrow&
(\bf{3}_{S},\bf{3}_{S})_{\rm{0}} \oplus
(\bf{3}_{S},\bf{2})_{\rm{-1}} \oplus (\bf{3}_{S},\bf{1})_{\rm{-2}} \\
(\bf{1}_{A},\bf{3}_{A}) &\rightarrow&
(\bf{1}_{A},\bf{1}_{A})_{\rm 0} \oplus (\bf{1}_{A},\bf{2})_{\rm -1},
\end{eqnarray} 
where subscripts $0,-1,-2$ denote the strangeness associated with each irrep.
Each of these irreps has one-to-one correspondence 
to the physical diquark state in singly heavy baryons,
as listed in the second last column of Table \ref{baryons}.


As for the light staggered quarks having four tastes
with degenerate mass,
the above $SU(3)_{F}$ flavor symmetry is extended to
$SU(12)_{f}$ flavor-taste symmetry \cite{Bailey}. 
Correspondingly, the staggered diquarks belong to the symmetric irreps
of $SU(24)$ which has the following decomposition,
\begin{eqnarray}
SU(24) &\supset& SU(2)_{S} \times SU(12)_{f} \nonumber \\
\bf{300}_{S} &\rightarrow& (\bf{3}_{S},\bf{78}_{S}) \oplus (\bf{1}_{A},\bf{66}_{A}).
\label{SU(24)}
\end{eqnarray} 
For $2+1$ flavor staggered quarks, 
the $SU(12)_{f}$ flavor-taste symmetry group is broken to
$SU(8)_{x,y}\times SU(4)_{z}$,
where $SU(8)_{x,y}$ denotes the symmetry group for two light valence quarks
while $SU(4)_{z}$ the one for a strange valence quark. 
The decomposition of $SU(12)_{f}$ into $SU(8)_{xy}\times SU(4)_{z}$ gives,
\begin{eqnarray}
SU(2)_{S}\times SU(12)_{f} &\supset& SU(2)_{S}\times 
SU(8)_{x,y} \times SU(4)_{z} \nonumber \\
({\bf{3}_{S}}, {\bf{78}_{S}}) &\rightarrow& 
({\bf{3}_{S}}, {\bf{36}_{S}},{\mathbf{1}})_{0} \oplus
({\bf{3}_{S}}, {\bf{8}},{\bf{4}})_{-1} \oplus 
({\bf{3}_{S}}, {\bf{1}},{\bf{10}_{S}})_{-2}, \label{spin_triplet} \\ 
({\bf{1}_{A}}, {\bf{66}_{A}}) &\rightarrow& 
({\bf{1}_{A}}, {\bf{28}_{A}},{\mathbf{1}})_{0} \oplus
({\bf{1}_{A}}, {\bf{8}},{\mathbf{4}})_{-1} \oplus 
({\bf{1}_{A}}, {\bf{1}},{\mathbf{6}_{A}})_{-2}, \label{spin_singlet}
\end{eqnarray}
We assume in the continuum limit that the taste symmetry 
restores and all the four tastes become equivalent.
Then we see 
all the irreps except $({\bf{1}_{A}},{\bf{1}},{\bf{6}_{A}})_{-2}$
are to be degenerate with physical diquarks under this assumption.
We list the staggered irreps for the physical diquarks 
in the last column of Table \ref{baryons}.
The strangeness $-2$ spin singlet diquark state
$({\bf{1}_{A}},{\bf{1}},{\bf{6}_{A}})_{-2}$ in (\ref{spin_singlet})
does not
correspond to any physical state in continuum limit.

In order to make contact with the lattice symmetry group,
we further decompose the physical states
into $SU(2)_{S}\times SU(2)_{I}\times SU(4)_{T}$
as follows,
\begin{eqnarray}
SU(2)_{S}\times SU(8)_{x,y}\times SU(4)_{z} &\supset& SU(2)_{S}\times 
SU(2)_{I} \times SU(4)_{T} \nonumber \\
\Sigma_{Q}^{(*)}
: \hspace{7pt} 
(\bf{3}_{S}, \bf{36}_{S},\bf{1})_{\rm 0} 
&\rightarrow & (\bf{3}_{S},\bf{3}_{S},\bf{10}_{S})_{\rm 0} \oplus
(\bf{3}_{S},\bf{1}_{A},\bf{6}_{A})_{\rm 0} \label{Sigma_2}\\
\Xi'^{(*)}_{Q}
:  \hspace{13pt}
(\bf{3}_{S},\bf{8},\bf{4})_{\rm -1}
&\rightarrow & (\bf{3}_{S},\bf{2},\bf{10}_{S})_{\rm -1} \oplus
(\bf{3}_{S},\bf{2},\bf{6}_{A})_{\rm -1} \\
\Omega_{Q}^{(*)}
: (\bf{3}_{S},\bf{1},\bf{10}_{S})_{\rm -2} 
&\rightarrow & (\bf{3}_{S},\bf{1},\bf{10}_{S})_{\rm -2} \\
\Lambda_{Q} : \hspace{3pt}
(\bf{1}_{A},\bf{28}_{A},\bf{1})_{\rm 0} 
&\rightarrow & (\bf{1}_{A},\bf{1}_{A},\bf{10}_{S})_{\rm 0} \oplus
(\bf{1}_{A},\bf{3}_{S},\bf{6}_{A})_{\rm 0}\\
\Xi_{Q} : \hspace{11pt}
(\bf{1}_{A},\bf{8},\bf{4})_{\rm -1} 
&\rightarrow & (\bf{1}_{A},\bf{2},\bf{10}_{S})_{\rm -1} \oplus
(\bf{1}_{A},\bf{2},\bf{6}_{A})_{\rm -1}. \label{Xi_2}
\end{eqnarray}
The main goal of this article is to construct the lattice staggered diquark
operators categorized into the physical irreps given in 
the right hand sides of
(\ref{Sigma_2})-(\ref{Xi_2}).

\section{Staggered Diquarks on the Lattice}

The symmetry group of staggered fermion action
on Euclidean lattice was first elaborated in \cite{Doel-Smit,G-S_self_energy}
and successively applied to classifying staggered 
baryons as well as mesons \cite{G-S,Golterman,Smit}.
The important symmetries of staggered fermions 
in our study are
$90^{\circ}$ 
rotations $R^{(\rho\sigma)}$, shift transformations $S_{\mu}$, 
space inversion $I_{s}$.
Since the shift operations $S_{\mu}$ contain
taste matrices,
pure translations $T_{\mu}$ may be represented by the square of $S_{\mu}$,
$T_{\mu}=S_{\mu}^{2}$.
Discrete taste transformations in Hilbert space are readily defined
by 
$\Xi_{\mu}\equiv S_{\mu}T^{-\frac{1}{2}}_{\mu}. \label{Xi_mu} $
The $\Xi_{\mu}$ generate 32 element Clifford group which 
is isomorphic to the discrete subgroup of $SU(4)_{T}$ in the continuum spacetime.
Since the space inversion contains a taste transformation $\Xi_{4}$,
the parity should be defined by 
$ P=\Xi_{4}I_{s} $. 
Note that the parity is non-locally defined in time direction 
since $\Xi_{4}$ is non-local in time.
For the purpose of spectroscopy, 
we are particularly interested in a symmetry group generated by 
the transformations which are local in time and
commuting with $T_{4}$.
Such a group is called geometrical time slice group ($GTS$)
which is given by
\begin{eqnarray}
GTS &=& G(R^{(kl)},\Xi_{m},I_{s}),
\end{eqnarray}
where $k,l, m = 1\sim 3$ \cite{G-S,Golterman,Smit}.
The defining representation of $GTS$ is 
given by the staggered quark fields
projected on zero spatial momentum.
It is an eight dimensional representation denoted as $\bf{8}$.
The anti-staggered quark fields 
also belong to the representation $\bf{8}$.
The $GTS$ representation of
staggered diquark is accordingly expressed by $\mathbf{8}\times\mathbf{8}$.
The decomposition of $\mathbf{8}\times\mathbf{8}$
into the bosonic irreps is given in \cite{Golterman},
\begin{eqnarray}
\mathbf{8}\times\mathbf{8} &=& 
\sum_{\sigma_{s}=\pm 1, \sigma_{123}=\pm 1}
\{
\mathbf{1}^{\sigma_{s}\sigma_{123}}
+\mathbf{3}^{\sigma_{s}\sigma_{123}}
+\mathbf{3''}^{\sigma_{s}\sigma_{123}}
+\mathbf{3''''}^{\sigma_{s}\sigma_{123}}
+\mathbf{6}^{\sigma_{s}\sigma_{123}}
\},
\end{eqnarray}
where
$\bf{1}$, $\bf{3}$, $\bf{3''}$, $\bf{3''''}$
and $\bf{6}$ are representing the bosonic representations
of $GTS$ with $\sigma_{s}$, the eigenvalue of $I_{s}$
and $\sigma_{123}$, the eigenvalue of $D(\Xi_{1}\Xi_{2}\Xi_{3})$.

The irreducibly transforming diquark operators  
are listed in Table \ref{diquark_operators} and \ref{diquark_operators2}.
As in the meson case, all the irreps are categorized 
into four classes from 0 to 3,
depending on how far the two staggered quarks are displaced each other.
The third column of the tables gives the operator form
of the diquarks.
The fourth column
gives the corresponding $GTS$ irreps.  
The $\eta_{\mu}$ and $\zeta_{\mu}$ denote the sign
factors defined by
$\eta_{\mu}(x)=(-1)^{x_{1}+\cdots +x_{\mu-1}}$ 
and $\zeta_{\mu}(x)=(-1)^{x_{\mu +1}+\cdots +x_{4}}$, respectively,
while $\epsilon$ is defined as $\epsilon(x)=(-1)^{x_{1}+x_{2}+x_{3}+x_{4}}$.  
The $D_{k}$ represents the symmetric shift operators defined by
$D_{k}\phi({\bf x}) = \frac{1}{2}[\phi({\bf x} +{\bf a}_{k})
+\phi({\bf x} -{\bf a}_{k})]$.
For notational simplicity, the sum over $\mathbf{x}$,
the color and flavor indices are suppressed without any confusion.
For example, $\chi \eta_{k}D_{k} \chi$
stands for
$
\sum_{\mathbf{x}}\chi^{a}_{f_{1}}({\bf x},t) 
\eta_{k}(x) D_{k} \chi^{b}_{f_{2}}({\bf x},t)
$.
As far as the lattice symmetry group $GTS$ is concerned, 
each diquark operator is formally corresponding to the meson operator
given in \cite{Golterman} through replacing the leftmost $\chi$ by $\overline{\chi}$.
This is because the staggered quark and anti-quark 
belong to the same $GTS$ irrep for each color and flavor.
The $\sigma_{4}$ in the fifth column denotes the eigenvalue of $X_{4}$ 
with which the parity 
is given by $P=\sigma_{s}\sigma_{4}$.
The sixth column gives
the spin and taste matrices 
$\Gamma_{S}\otimes\Gamma_{T}$
which come into the diquark operators in the spin-taste basis,
$
\psi^{T}(C\Gamma_{S}\otimes (\Gamma_{T}C^{-1})^{T})\psi,
$
where the superscript $T$ denotes transpose
and $C$ denotes the charge conjugation matrix.
The presence of $C$ and $C^{-1}$
ensures
the covariant properties under the spin and taste 
rotations in the continuum limit.
Notice that
the assignment of $\Gamma_{S}\otimes \Gamma_{T}$ for 
each $GTS$ irrep is systematically different from the meson case, 
where the operators are given by 
$\overline{\psi}(\Gamma_{S}\otimes(\Gamma_{T})^{T})\psi$
in the spin-taste basis.


\begin{table}
\begin{center}
\renewcommand{\arraystretch}{1.1}
\renewcommand{\tabcolsep}{8pt}
\begin{scriptsize}
\begin{tabular}{ccccccccc}
\hline
class & No. & operator & $GTS$ ($\overline{\mathbf{r}}^{\sigma_{s}\sigma_{123}}$) &
$\sigma_{4}$ & $\Gamma_{S}\otimes \Gamma_{T}$  
& $J^{P}$  & order & $(SU(2)_{S},SU(4)_{T})$\\ \hline
$0$ & $1$ & $\chi\chi$ & $\mathbf{1}^{++}$ & $+$ & $\gamma_{5}\otimes\gamma_{5}$ 
& $0^{+}$ &   $1$ & $(\bf{1}_{A},\bf{6}_{A})$\\
& & &  & $-$ & $\gamma_{4}\otimes\gamma_{4}$ 
& $0^{-}$ & $p/E$\\ 
& $2$ & $\eta_{4}\zeta_{4}\chi\chi$ 
& $\mathbf{1}^{+-}$ & $+$ & $\gamma_{4}\gamma_{5}\otimes\gamma_{4}\gamma_{5}$ 
 & $0^{+}$ & $1$ & $(\bf{1}_{A},\bf{6}_{A})$\\
&&&  & $-$ & $1\otimes1$ 
& $0^{-}$ & $p/E$\\ 
& $3$ & $\eta_{k}\epsilon\zeta_{k}\chi\chi$ 
& $\mathbf{3''''}^{+-}$ & $+$ & $\gamma_{k}\otimes\gamma_{k}$ 
& $1^{+}$ &  $1$ 
& $(\bf{3}_{S},\bf{10}_{S})$\\
&&&  & $-$ & $\gamma_{l}\gamma_{m}\otimes\gamma_{l}\gamma_{m}$ 
&$1^{-}$ &  $p/E$\\ 
&$4$ &$\eta_{4}\zeta_{4}\eta_{k}\epsilon\zeta_{k}\chi\chi$ 
& $\mathbf{3''''}^{++}$ & $+$ & $\gamma_{k}\gamma_{4}\otimes\gamma_{k}\gamma_{4}$ 
&$1^{+}$ &  $1$
& $(\bf{3}_{S},\bf{10}_{S})$\\
&&&  & $-$ & $\gamma_{k}\gamma_{5}\otimes\gamma_{k}\gamma_{5}$ 
& $1^{-}$ & $p/E$\\[10pt] 
$1$ & $5$ &
$\chi\eta_{k}D_{k}\chi$ 
& $\mathbf{3}^{-+}$ & $+$ & $\gamma_{k}\gamma_{5}\otimes\gamma_{5}$ 
& $1^{-}$ & $p/E$ \\
&&&  & $-$ & $\gamma_{k}\gamma_{4}\otimes\gamma_{4}$ 
&$1^{+}$ & $1$
& $(\bf{3}_{S},\bf{10}_{S})$\\ 
&$6$ &$\eta_{4}\zeta_{4}\chi\eta_{k}D_{k}\chi$ 
& $\mathbf{3}^{--}$ & $+$ & $\gamma_{l}\gamma_{m}\otimes\gamma_{4}\gamma_{5}$ 
& $1^{-}$ &  $p/E$\\
&&&  & $-$ & $\gamma_{k}\otimes 1$ 
&$1^{+}$ &  $1$
& $(\bf{3}_{S},\bf{6}_{A})$ \\ 
&$7$ &$\chi\epsilon\zeta_{k}D_{k}\chi$ 
& $\mathbf{3}''^{--}$ & $+$ & $1\otimes\gamma_{k}$ 
& $0^{-}$ &  $p/E$\\
&&&  & $-$ & $\gamma_{4}\gamma_{5}\otimes \gamma_{l}\gamma_{m}$ 
& $0^{+}$ &  $1$
& $(\bf{1}_{A},\bf{10}_{S})$\\ 
&$8$&$\eta_{4}\zeta_{4}\chi\epsilon\zeta_{k}D_{k}\chi$ 
& $\mathbf{3}''^{-+}$ & $+$ & $\gamma_{4}\otimes\gamma_{k}\gamma_{4}$ 
&$0^{-}$ & $p/E$\\
&&&  & $-$ & $\gamma_{5}\otimes \gamma_{k}\gamma_{5}$ 
& $0^{+}$ & $1$ & $(\bf{1}_{A},\bf{6}_{A})$\\ 
&$9$ &$\eta_{k}\epsilon\zeta_{k}\chi\eta_{l}D_{l}\chi$ 
& $\mathbf{6}^{--}$ & $+$ & $\gamma_{k}\gamma_{l}\otimes\gamma_{k}$ 
& $1^{-}$ &  $p/E$\\
&&&  & $-$ & $\gamma_{m}\otimes \gamma_{l}\gamma_{m}$ 
& $1^{+}$ & $1$
& $(\bf{3}_{S},\bf{10}_{S})$\\ 
&$10$ & $\eta_{4}\zeta_{4}\eta_{k}\epsilon\zeta_{k}\chi\eta_{l}D_{l}\chi$ 
& $\mathbf{6}^{-+}$ & $+$ & $\gamma_{m}\gamma_{5}\otimes\gamma_{k}\gamma_{4}$ 
& $1^{-}$ &  $p/E$\\
&&&  & $-$ & $\gamma_{m}\gamma_{4}\otimes \gamma_{k}\gamma_{5}$ 
& $1^{+}$ &  $1$ & $(\bf{3}_{S},\bf{6}_{A})$  \\ \hline
\end{tabular}
\end{scriptsize}
\caption{$GTS$ irrep., $\sigma_{4}$, $\Gamma_{S}\otimes \Gamma_{T}$
and continuum states for staggered diquark operators up to class 1.
($k,l,m = 1\sim 3,\ k\neq l \neq m \neq k$).
The summation over $\bf{x}$, flavor and color indices are omitted.}
\label{diquark_operators}
\end{center}
\end{table}

\begin{table}
\begin{center}
\renewcommand{\arraystretch}{1.1}
\renewcommand{\tabcolsep}{5pt}
\begin{scriptsize}
\begin{tabular}{ccccccccc}
\hline
class & No. & operator & $GTS$ ($\overline{\mathbf{r}}^{\sigma_{s}\sigma_{123}}$) &
$\sigma_{4}$ & $\Gamma_{S}\otimes \Gamma_{T}$  
& $J^{P}$ &  order & $(SU(2)_{S},SU(4)_{T})$\\ \hline
$2$ &$11$& $\chi\eta_{k}D_{k}\{\eta_{l}D_{l}\chi\}$ 
& $\mathbf{3}^{++}$ & $+$ & $\gamma_{m}\gamma_{4}\otimes\gamma_{5}$ 
& $1^{+}$ &  $1 $ & $(\bf{3}_{S},\bf{6}_{A})$\\
& & && $-$ & $\gamma_{m}\gamma_{5}\otimes\gamma_{4}$
& $1^{-}$ &  $p/E$ \\
 & $12$&$\eta_{4}\zeta_{4}\chi\eta_{k}D_{k}\{\eta_{l}D_{l}\chi\}$ 
& $\mathbf{3}^{+-}$ & $+$ & $\gamma_{m}\otimes\gamma_{4}\gamma_{5}$ 
& $1^{+}$ &  $1$ & $(\bf{3}_{S},\bf{6}_{A})$\\
& && & $-$ & $\gamma_{k}\gamma_{l}\otimes 1$
& $1^{-}$ &  $p/E$\\
 & $13$ &$\chi\zeta_{k}D_{k}\{\zeta_{l}D_{l}\chi\}$ 
& $\mathbf{3}''^{++}$ & $+$ & $\gamma_{5}\otimes\gamma_{m}\gamma_{4}$ 
& $0^{+}$ &  $1$
& $(\bf{1}_{A},\bf{10}_{S})$ \\
& & & &$-$ & $\gamma_{4}\otimes \gamma_{m}\gamma_{5}$
& $0^{-}$ &  $p/E$\\
 & $14$ &$\eta_{4}\zeta_{4}\chi\zeta_{k}D_{k}\{\zeta_{l}D_{l}\chi\}$ 
& $\mathbf{3}''^{+-}$ & $+$ & $\gamma_{4}\gamma_{5}\otimes\gamma_{m}$ 
& $0^{+}$ &  $1$
& $(\mathbf{1}_{A},\mathbf{10}_{S})$\\
& & && $-$ & $1 \otimes \gamma_{k}\gamma_{l}$
& $0^{-}$ &  $p/E$\\
 & $15$ &$\eta_{m}\zeta_{m}\chi\eta_{k}D_{k}\{\zeta_{l}D_{l}\chi\}$ 
& $\mathbf{6}^{++}$ & $+$ & $\gamma_{l}\gamma_{4}\otimes\gamma_{k}\gamma_{4}$ 
& $1^{+}$ &  $1$ 
& $(\bf{3}_{S},\bf{10}_{S})$\\
& & && $-$ & $\gamma_{l}\gamma_{5} \otimes \gamma_{k}\gamma_{5}$
& $1^{-}$ &  $p/E$\\
 & $16$ &$\eta_{4}\zeta_{4}\eta_{m}\zeta_{m}\chi\eta_{k}D_{k}\{\zeta_{l}D_{l}\chi\}$ 
& $\mathbf{6}^{+-}$ & $+$ & $\gamma_{l}\otimes\gamma_{k}$ 
& $1^{+}$ &  $1$
& $(\bf{3}_{S},\bf{10}_{S})$\\
& & & &$-$ & $\gamma_{k}\gamma_{m} \otimes \gamma_{l}\gamma_{m}$
& $1^{-}$ &  $p/E$\\[10pt]
$3$ &$17$ & $\chi\eta_{1}D_{1}\{\eta_{2}D_{2}\{\eta_{3}D_{3}\chi\}\}$ 
& $\mathbf{1}^{-+}$ & $+$ & $\gamma_{4}\otimes\gamma_{5}$ 
& $0^{-}$ &  $p/E$\\
& & & & $-$ & $\gamma_{5} \otimes \gamma_{4}$
& $0^{+}$ &  $1$
&  $(\bf{1}_{A},\bf{10}_{S})$\\
 & $18$ &$\eta_{4}\zeta_{4}\chi\eta_{1}D_{1}\{\eta_{2}D_{2}\{\eta_{3}D_{3}\chi\}\}$ 
& $\mathbf{1}^{--}$ & $+$ & $ 1\otimes\gamma_{4}\gamma_{5}$ 
& $0^{-}$ &  $p/E$\\
& & & &$-$ & $\gamma_{4}\gamma_{5} \otimes 1$
& $0^{+}$ &  $1$ & $(\bf{1}_{A},\bf{6}_{A})$\\
 & $19$ &$\eta_{k}\epsilon\zeta_{k}\chi\eta_{1}D_{1}
\{\eta_{2}D_{2}\{\eta_{3}D_{3}\chi\}\}$ 
& $\mathbf{3}''''^{--}$ & $+$ & $ \gamma_{l}\gamma_{m}\otimes \gamma_{k}$ 
& $1^{-}$ &  $p/E$\\
& & & &$-$ & $\gamma_{k}\otimes \gamma_{l}\gamma_{m}$
& $1^{+}$ & $1$
& $(\bf{3}_{S},\bf{10}_{S})$\\
 & $20$ &$\eta_{4}\zeta_{4}\eta_{k}\epsilon\zeta_{k}
\chi\eta_{1}D_{1}\{\eta_{2}D_{2}\{\eta_{3}D_{3}\chi\}\}$ 
& $\mathbf{3}''''^{-+}$ & $+$ & $ \gamma_{k}\gamma_{5}\otimes \gamma_{k}\gamma_{4}$ 
& $1^{-}$ &  $p/E$\\
& & && $-$ & $\gamma_{k}\gamma_{4}\otimes \gamma_{k}\gamma_{5}$
& $1^{+}$ &  $1$ & $(\bf{3}_{S},\bf{6}_{A})$ \\
\hline
\end{tabular}
\end{scriptsize}
\caption{$GTS$ irrep., $\sigma_{4}$, $\Gamma_{S}\otimes \Gamma_{T}$
and continuum states for staggered diquark operators class 2 and 3.
($k,l,m = 1\sim 3,\ k\neq l \neq m \neq k$).
The summation over $\bf{x}$, flavor and color indices are omitted.}
\label{diquark_operators2}
\end{center}
\end{table}


\section{Connection between lattice and continuum irreps}

Consulting the relations between lattice $\overline{RF}$ irreps and
continuum spin irreps given in \cite{Golterman}
and assuming that the ground states of lattice irreps correspond to 
the lowest possible spin in the continuum limit,
one could make an assignment of spin and parity $J^{P}$
for each $GTS$ irrep.
See the seventh column of
Tables \ref{diquark_operators} and \ref{diquark_operators2}. 
One also see that
the combinations, 
$C\Gamma_{S} = C\gamma_{k}, 
C\gamma_{k}\gamma_{4},
C\gamma_{4}\gamma_{5},
C\gamma_{5}, 
$
generate $upper \times upper$ 
products of the Dirac spinors 
in Dirac representation 
for each taste
and then give rise to 
$\mathcal{O}(1)$ contributions,
while the combinations,
$
C\Gamma_{S} = C, 
C\gamma_{4},
C\gamma_{k}\gamma_{l},
C\gamma_{k}\gamma_{5}, 
$
generate $upper \times lower$ products,
so that they are suppressed by $\mathcal{O}(p/E)$
in the non-relativistic limit.
See the second last column of Table \ref{diquark_operators} and
\ref{diquark_operators2}.
An important notice here is
that only the positive parity states 
survive in the non-relativistic limit,
which is in accordance with the property of
physical diquarks.
As for the $SU(4)_{T}$ irreps, one sees that 
the combinations,
$
\Gamma_{T}C^{-1} = \gamma_{k}C^{-1}, 
\gamma_{4}C^{-1},
\gamma_{k}\gamma_{l}C^{-1},
\gamma_{k}\gamma_{4}C^{-1},
$
are symmetric 
so that they 
altogether belong to $\bf{10}_{S}$ irrep of $SU(4)_{T}$,
while the anti-symmetric 
six combinations
$
\Gamma_{T}C^{-1} = C^{-1},
\gamma_{k}\gamma_{5}C^{-1},
\gamma_{4}\gamma_{5}C^{-1},
\gamma_{5}C^{-1},
$
belong to
$\bf{6}_{A}$ irrep of $SU(4)_{T}$.
The assignments of non-relativistic $SU(2)_{S}\times SU(4)_{T}$ 
for the lattice irreps
are readily given for the $\mathcal{O}(1)$ operators. 
They are listed  in the last column of 
the tables.
The final step 
is to take into account the $2+1$ flavor symmetry,
which could be done in a straightforward manner.
The decomposition of continuum spin, $2+1$ flavor and taste 
symmetry group  
into the lattice symmetry group is given by,
\begin{eqnarray}
SU(2)_{S}\times SU(2)_{I}\times SU(4)_{T}
&\supset& SU(2)_{I}\times GTS.
\end{eqnarray}
In Table \ref{result}, we list all the lattice diquark operators
which are local in time
and 
categorize them into
each continuum irrep $(SU(2)_{S},SU(2)_{I},SU(4)_{T})_{Z}$
previously 
given in (\ref{Sigma_2})-(\ref{Xi_2}).

\section{Summary}

Continuum and lattice irreps of staggered diquarks with
$SU(4)$ taste symmetry in $2+1$ flavors were studied.
We have started from the $SU(24)$ symmetry group which is 
the $SU(4)$ taste extension of ordinary $SU(6)$
non-relativistic quark model.
This procedure has been also taken
in the study of staggered baryon classifications 
\cite{Bailey}.  
As for the lattice representations,
we have consulted the lattice symmetry group of staggered fermion action
elaborated in \cite{Doel-Smit,G-S_self_energy}.
Although the irreps of lattice symmetry group $GTS$
cannot have any definite parity, 
we have explicitly shown 
that only the positive parity state
contributes in the non-relativistic limit,
which is in accordance with the property of physical diquarks.

\section*{Acknowledgments}

We would like to thank S. Basak, C. Bernard and
C. DeTar for useful discussions and comments.
We thank J. Bailey for important comments on the physical states.
This work has been supported by U.S. Department of Energy, Grant No.
FG02-91ER 40661.


\begin{table}
\renewcommand{\arraystretch}{1.05}
\renewcommand{\tabcolsep}{1pt}
\begin{scriptsize}
\begin{tabular}{|c|c|c|c|}
\hline
No. 
& $\Sigma_{Q}^{(*)}
: (\bf{3}_{S},\bf{3}_{S},\bf{10}_{S})_{\rm 0}$
& $\Xi'^{(*)}_{Q}
: (\bf{3}_{S},\bf{2},\bf{10}_{S})_{\rm -1}$
& $\Omega_{Q}^{(*)}
: (\bf{3}_{S},\bf{1},\bf{10}_{S})_{\rm -2}$ \\ \hline
$3$ & $\eta_{k}\epsilon\zeta_{k}ll$ 
& $\eta_{k}\epsilon\zeta_{k}ls+\eta_{k}\epsilon\zeta_{k}sl$
& $\eta_{k}\epsilon\zeta_{k}ss$ \\
 $4$ & $\eta_{4}\zeta_{4}\eta_{k}\epsilon\zeta_{k}ll$ 
& $\eta_{4}\zeta_{4}\eta_{k}\epsilon\zeta_{k}ls+
\eta_{4}\zeta_{4}\eta_{k}\epsilon\zeta_{k}sl$
& $\eta_{4}\zeta_{4}\eta_{k}\epsilon\zeta_{k}ss$ \\ 
$5$ & $l\eta_{k}D_{k}l$ 
& $l\eta_{k}D_{k}s + s\eta_{k}D_{k}l$
& $s\eta_{k}D_{k}s $ \\
$9$ & $\eta_{k}\epsilon\zeta_{k}l\eta_{l}D_{l}l$
& $\eta_{k}\epsilon\zeta_{k}l\eta_{l}D_{l}s 
+\eta_{k}\epsilon\zeta_{k}s\eta_{l}D_{l}l$ 
& $\eta_{k}\epsilon\zeta_{k}s\eta_{l}D_{l}s$ \\ 
$15$ & $\eta_{m}\zeta_{m}l\eta_{k}D_{k}\{\zeta_{l}D_{l}l\}$ 
& $\eta_{m}\zeta_{m}l\eta_{k}D_{k}\{\zeta_{l}D_{l}s\}
+ \eta_{m}\zeta_{m}s\eta_{k}D_{k}\{\zeta_{l}D_{l}l\}$
& $\eta_{m}\zeta_{m}s\eta_{k}D_{k}\{\zeta_{l}D_{l}s\}$ \\
$16$ & $\eta_{4}\zeta_{4}\eta_{m}\zeta_{m}l\eta_{k}D_{k}\{\zeta_{l}D_{l}l\}$ 
& $\eta_{4}\zeta_{4}\eta_{m}\zeta_{m}l\eta_{k}D_{k}\{\zeta_{l}D_{l}s\}
+ \eta_{4}\zeta_{4}\eta_{m}\zeta_{m}s\eta_{k}D_{k}\{\zeta_{l}D_{l}l\}$
& $\eta_{4}\zeta_{4}\eta_{m}\zeta_{m}s\eta_{k}D_{k}\{\zeta_{l}D_{l}s\}$ \\
$19$ 
& $\eta_{k}\epsilon\zeta_{k}l\eta_{1}D_{1}\{\eta_{2}D_{2}\{\eta_{3}D_{3}l\}\}$
& $\eta_{k}\epsilon\zeta_{k}l\eta_{1}D_{1}\{\eta_{2}D_{2}\{\eta_{3}D_{3}s\}\}
+ \eta_{k}\epsilon\zeta_{k}s\eta_{1}D_{1}\{\eta_{2}D_{2}\{\eta_{3}D_{3}l\}\}$
& $\eta_{k}\epsilon\zeta_{k}s\eta_{1}D_{1}\{\eta_{2}D_{2}\{\eta_{3}D_{3}s\}\}$
\\ \hline
\end{tabular}
\begin{tabular}{|c|c|c|}
\hline
No. 
& $\Sigma_{Q}^{(*)}
: (\bf{3}_{S},\bf{1}_{A},\bf{6}_{A})_{\rm 0}$ 
& $\Xi'^{(*)}_{Q}
: (\bf{3}_{S},\bf{2},\bf{6}_{A})_{\rm -1}$ \\ \hline
$6$ & $\eta_{4}\zeta_{4}l_{1}\eta_{k}D_{k}l_{2} 
- \eta_{4}\zeta_{4}l_{2}\eta_{k}D_{k}l_{1}$ 
& $\eta_{4}\zeta_{4}l\eta_{k}D_{k}s 
- \eta_{4}\zeta_{4}s\eta_{k}D_{k}l$ \\
$10$ & $\eta_{4}\zeta_{4}\eta_{k}\epsilon\zeta_{k}l_{1}\eta_{l}D_{l}l_{2}
- \eta_{4}\zeta_{4}\eta_{k}\epsilon\zeta_{k}l_{2}\eta_{l}D_{l}l_{1}$ 
& $\eta_{4}\zeta_{4}\eta_{k}\epsilon\zeta_{k}l\eta_{l}D_{l}s
- \eta_{4}\zeta_{4}\eta_{k}\epsilon\zeta_{k}s\eta_{l}D_{l}l$ \\ 
$11$ 
& $l_{1}\eta_{k}D_{k}\{\eta_{l}D_{l}l_{2}\}
- l_{2}\eta_{k}D_{k}\{\eta_{l}D_{l}l_{1}\}$
& $l\eta_{k}D_{k}\{\eta_{l}D_{l}s\}
- s\eta_{k}D_{k}\{\eta_{l}D_{l}l\}$ \\
$12$ 
& $\eta_{4}\zeta_{4}l_{1}\eta_{k}D_{k}\{\eta_{l}D_{l}l_{2}\}
- \eta_{4}\zeta_{4}l_{2}\eta_{k}D_{k}\{\eta_{l}D_{l}l_{1}\}$
& $\eta_{4}\zeta_{4}l\eta_{k}D_{k}\{\eta_{l}D_{l}s\}
- \eta_{4}\zeta_{4}s\eta_{k}D_{k}\{\eta_{l}D_{l}l\}$ \\
$20$ 
& $\eta_{4}\zeta_{4}\eta_{k}\epsilon\zeta_{k}l_{1}
\eta_{1}D_{1}\{\eta_{2}D_{2}\{\eta_{3}D_{3}l_{2}\}\}
-\eta_{4}\zeta_{4}\eta_{k}\epsilon\zeta_{k}l_{2}
\eta_{1}D_{1}\{\eta_{2}D_{2}\{\eta_{3}D_{3}l_{1}\}\}$ 
& $\eta_{4}\zeta_{4}\eta_{k}\epsilon\zeta_{k}l
\eta_{1}D_{1}\{\eta_{2}D_{2}\{\eta_{3}D_{3}s\}\}
-\eta_{4}\zeta_{4}\eta_{k}\epsilon\zeta_{k}s
\eta_{1}D_{1}\{\eta_{2}D_{2}\{\eta_{3}D_{3}l\}\}$
 \\ \hline
\end{tabular}
\begin{tabular}{|c|c|c|}
\hline
No. 
& $\Lambda_{Q} : (\bf{1}_{A},\bf{1}_{A},\bf{10}_{S})_{\rm 0}$ 
& $\Xi_{Q} : (\bf{1}_{A},\bf{2},\bf{10}_{S})_{\rm -1}$ \\ \hline
$7$ & $l_{1}\epsilon\zeta_{k}D_{k}l_{2} - l_{2}\epsilon\zeta_{k}D_{k}l_{1}$ 
& $l\epsilon\zeta_{k}D_{k}s - s\epsilon\zeta_{k}D_{k}l$ \\
$13$ 
& $l_{1}\zeta_{k}D_{k}\{\zeta_{l}D_{l}l_{2}\}
- l_{2}\zeta_{k}D_{k}\{\zeta_{l}D_{l}l_{1}\}$
& $l\zeta_{k}D_{k}\{\zeta_{l}D_{l}s\}
- s\zeta_{k}D_{k}\{\zeta_{l}D_{l}l\}$ \\
$14$ 
& $\eta_{4}\zeta_{4}l_{1}\zeta_{k}D_{k}\{\zeta_{l}D_{l}l_{2}\}
- \eta_{4}\zeta_{4}l_{2}\zeta_{k}D_{k}\{\zeta_{l}D_{l}l_{1}\}$
& $\eta_{4}\zeta_{4}l\zeta_{k}D_{k}\{\zeta_{l}D_{l}s\}
- \eta_{4}\zeta_{4}s\zeta_{k}D_{k}\{\zeta_{l}D_{l}l\}$ \\
$17$ 
& $l_{1}\eta_{1}D_{1}\{\eta_{2}D_{2}\{\eta_{3}D_{3}l_{2}\}\}
- l_{2}\eta_{1}D_{1}\{\eta_{2}D_{2}\{\eta_{3}D_{3}l_{1}\}\} $
& $l\eta_{1}D_{1}\{\eta_{2}D_{2}\{\eta_{3}D_{3}s\}\}
- s\eta_{1}D_{1}\{\eta_{2}D_{2}\{\eta_{3}D_{3}l\}\} $
\\ \hline
\end{tabular}
\begin{tabular}{|c|c|c|}
\hline
No. 
& $\Lambda_{Q} : (\bf{1}_{A},\bf{3}_{S},\bf{6}_{A})_{\rm 0}$ 
& $\Xi_{Q} : (\bf{1}_{A},\bf{2},\bf{6}_{A})_{\rm -1}$ \\ \hline
$1$ & $ll$ & $ls+sl$ \\
$2$ & $\eta_{4}\zeta_{4}ll$ & $\eta_{4}\zeta_{4}ls+\eta_{4}\zeta_{4}sl$ \\
$8$ & $\eta_{4}\zeta_{4}l\epsilon\zeta_{k}D_{k}l$ 
& $\eta_{4}\zeta_{4}l\epsilon\zeta_{k}D_{k}s
+ \eta_{4}\zeta_{4}s\epsilon\zeta_{k}D_{k}l$ \\
$18$ 
& $\eta_{4}\zeta_{4}l\eta_{1}D_{1}\{\eta_{2}D_{2}\{\eta_{3}D_{3}l\}\}$
& $\eta_{4}\zeta_{4}l\eta_{1}D_{1}\{\eta_{2}D_{2}\{\eta_{3}D_{3}s\}\}
+ \eta_{4}\zeta_{4}s\eta_{1}D_{1}\{\eta_{2}D_{2}\{\eta_{3}D_{3}l\}\}$
\\ \hline
\end{tabular}
\end{scriptsize}
\caption{Lattice staggered diquark operators categorized into the continuum 
irreps $(SU(2)_{S},SU(2)_{I},SU(4)_{T})_{Z}$ : 
The $l$ and $s$ denote light and strange quark, respectively.
No. indicates the operator number in Table 
2 and 3.
The summation over $\bf{x}$ and color indices are omitted.}
\label{result}
\end{table}



\begin{thebibliography}{99}


\bibitem{Bowler}
K.C.~Bowler {\it et al.} (UKQCD Collaboration),
Phys. Rev. {\bf D54} (1996) 3619.

\bibitem{Flynn}
J.~M.~Flynn, F.~Mescia and A.~S.~B.~Tariq  (UKQCD Collaboration),
JHEP {\bf 0307}, 066 (2003)

\bibitem{Mathur}
N.~Mathur {\it et al.},
Phys. Rev. {\bf D66} (2002) 014502.

\bibitem{Woloshyn}
R.M.~Woloshyn,
Phys. Lett. {\bf B476} (2000) 309.

\bibitem{G-T}
S.~Gottlieb and S.~Tamhankar,
Nucl. Phys. Proc. Suppl. {\bf 119} (2003) 644.

\bibitem{Khan}
A.~Ali Khan {\it et al.},
Phys. Rev. {\bf D62} (2000) 054505.

\bibitem{Chiu}
T.W.~Chiu and T.H.Hsieh,
Nucl. Phys. {\bf A755} (2005) 471c.


\bibitem{N-G}
S.~Gottlieb and H.~Na, 
PoS {\bf LAT2006}, 191 (2006)
[{\tt hep-lat/0610009}].


\bibitem{MILC}
C.~Aubin {\it et al.}(MILC Collaboration),
Phys. Rev. {\bf D70} (2004) 114501.


\bibitem{GNN}
S.~Gottlieb, H.~Na and K.~Nagata,
  arXiv:0707.3537 [hep-lat].



\bibitem{Bailey}
J.~Bailey, 
Phys.\ Rev.\  {\bf D75}, 114505 (2007).


\bibitem{G-S}
M.F.L.~Golterman and J.~Smit,
Nucl. Phys. {\bf B255} (1985) 328.

\bibitem{Golterman}
M.F.L.~Golterman,
Nucl. Phys. {\bf B273} (1986) 663.

\bibitem{Smit}
J.~Smit,
``Hadron Operators for Staggered Fermions", \\
at Conf. Advances in Lattice Gauge Theory, 
Tallahassee, Fla., Apr 10-13, 1985. 

\bibitem{Doel-Smit}
C.P.~van den Doel and J.~Smit,
Nucl. Phys. {\bf B228} (1983) 122.

\bibitem{G-S_self_energy}
M.F.L.~Golterman and J.~Smit,
Nucl. Phys. {\bf B245} (1984) 61.


\bibitem{Gupta}
S.~Gupta,
Phys. Rev. {\bf D60} (1999) 094505. 






\end{thebibliography}
\end{document}